\documentclass[twocolumn,showpacs,preprintnumbers,amsmath,amssymb,pre]{revtex4-1}

\usepackage{graphicx}
\usepackage{dcolumn}
\usepackage{overpic}
\usepackage{float} 
\usepackage{bbm}

\begin{document}

\title{Topologically-enforced bifurcations in  superconducting circuits}

\author{ G. Engelhardt$^1$    }
\email{georg@itp.tu-berlin.de}

\author{ M. Benito$^2$}

\author{ G. Platero$^2$}

\author{ T. Brandes$^1$}

\affiliation{%
$^1$Institut f\"ur Theoretische Physik, Technische Universit\"at Berlin, Hardenbergstr. 36, 10623 Berlin, Germany \\
$^2$Instituto de Ciencia de Materials de Madrid, CSIC, 28049 Madrid, Spain}%

\date{\today}


\begin{abstract}
The relation of topological insulators and superconductors and the field of nonlinear dynamics is widely unexplored. To address this subject, we adopt the linear coupling geometry of the Su-Schrieffer-Heeger model, a paradigmatic example for a topological insulator,  and render it nonlinearly in the context of superconducting circuits. As a consequence, the system exhibits  topologically-enforced bifurcations as a function of the topological control parameter, which finally gives rise to chaotic dynamics, separating phases which exhibit  clear topological features.
\end{abstract}
\maketitle

\textit{Introduction}.
Topological insulators and superconductors have attracted much attention in  recent time. Prominent examples are the integer quantum-Hall effect, chiral edge bands or topologically-protected Majorana fermions~\cite{Thouless1982,Hasan2010,Bernevig2013}. These effects are thereby a consequence of a linear, but non-trivial band structure of noninteracting particles, so that they can also appear in  bosonic and even classical systems~\cite{Suesstrunk2015,Rechtsman2013,Hafezi2013,McHugh2016,Engelhardt2015,Peano2016a,Engelhardt2016,Peano2016}.

However, in actual physical systems nonlinearities are omnipresent, either desired or not. They give rise to outstanding and various effects as bifurcations, synchronization  and chaos   appearing  in different kinds of fields  reaching from cold atoms, biology,  chemistry to superconducting circuits~\cite{Tomkovivc2015,Baumann2010,Strogatz2014,Kautz1996}. For this reason it is interesting to ask about the relation of nonlinear dynamics and linear topological effects.

\begin{figure}
\centerline{\includegraphics[width=\columnwidth]{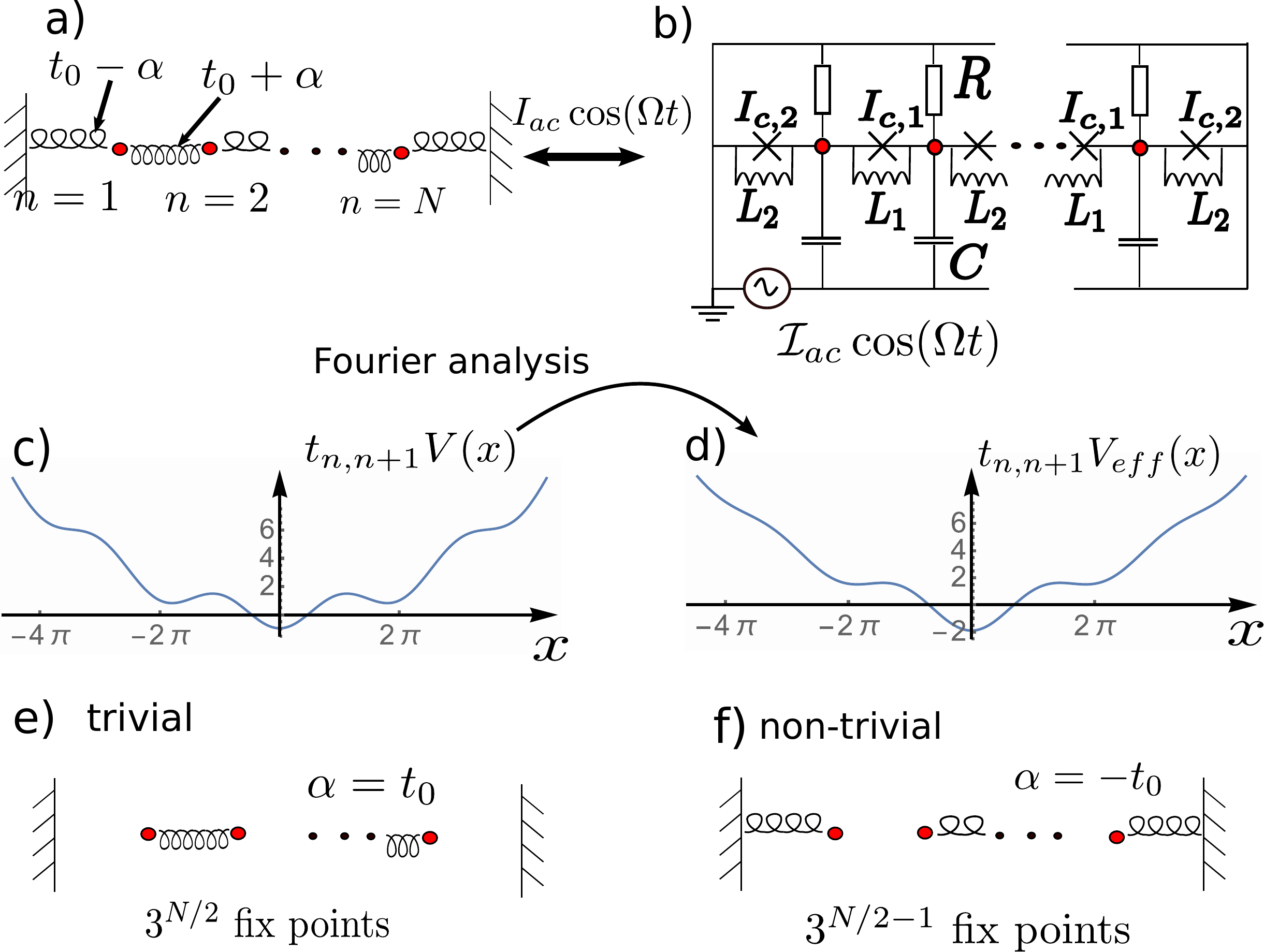}}
\caption{(a) Sketch of the system. The coupling geometry resembles the one of the SSH model with alternating coupling strength.  The system is subjected to external driving and dissipation (non sketched).  (b) Superconducting circuit giving rise to Eq.~\eqref{eq:EoM}. (c)  Nonlinear coupling  potential of the nodes. (d) Effective coupling potential appearing due to a Fourier analysis of Eq.~\eqref{eq:EoM}. (e) Topologically trivial limiting case $\alpha=t_0$, where the system consists of uncoupled dimers. (f) Topologically non-trivial limiting case $\alpha=-t_0$, where two uncoupled nodes at the boundary exist.}
\label{fig:sketch}
\end{figure}

 One of the simplest models exhibiting topological effects is the celebrated Su-Schrieffer-Heeger (SSH) model~\cite{Su1979,Asboth2016}, which features topologically-protected boundary excitation due to its coupling geometry as sketched in Fig.~\ref{fig:sketch}(a). Thereby, the topological effects can be explained using linear algebra. In this Letter, we propose a realization of the SSH model in superconducting circuits that allows one to study the impact of nonlinearites on topological properties in a controlled way, i.e., by using external (ac) driving (see Fig.~\ref{fig:sketch}(b)).

We demonstrate that the nonlinearly-rendered SSH model exhibits  topologically-enforced bifurcations which lead to  chaotic dynamics. Our analysis is based on an effective  coupling potential and refers to the number of fix points of two specific topologically distinct limiting cases which are depicted in Fig~\ref{fig:sketch}(e),(f).  Although referring here to a very specific model, our findings are relevant for all kind of lattice models with possible topological coupling geometry, where nonlinearities are so strong that 
bifurcations can occur, as in cold-atomic systems~\cite{Goldman2016,Morsch2006,Aidelsburger2013}, optomechanics~\cite{Purdy2013} or optics with nonlinear materials~\cite{Mookherjea2002,Eggleton2011,Dahdah2011}.

 In the  literature, the effect of nonlinearities due to interactions are mostly considered in the context of ground-state properties of topological systems~\cite{Gurarie2011}.  Another famous subject are fractional excitations close to the ground state~\cite{Laughlin1983,Tsui1982,Stormer1999,Grusdt2013}. Very recently, topological phase transitions induced by a combination of driving and nonlinearities have been investigated~\cite{Hadad2016}. Here, we follow a different approach by investigating the complex nonlinear dynamics, for which, in principal, the total phase space is relevant.

\textit{The system}. We consider  a one-dimensional system of $N$ nonlinearly coupled nodes as sketched in Fig.~\ref{fig:sketch}(a),(b). The equations of motion (EoM) determining the dynamics read
\begin{align}
	\ddot\phi_{n} &= t_{n-1,n}  f_\delta \left(\phi_{n-1}-\phi_{n} \right)+  t_{n,n+1}  f_\delta \left(\phi_{n+1}-\phi_{n} \right) \nonumber \\
	&-\mathcal R \dot \phi_{n}  +  I_{ac} \cos(\Omega t),
	\label{eq:EoM}
\end{align}
where the nonlinearity enters via the function
\begin{equation}
	f_\delta(x) = \left(1-\delta\right)x+ \delta \sin\left( x\right).
	\label{eq:couplingFunction}
\end{equation}
These EoM can be modeled by   a system of superconducting islands coupled by  inductively shunted  Josephson junctions as sketched in Fig.~\ref{fig:sketch}(b)~\cite{Supplementals,Manucharyan2009,Erguel2013,Koch2009,Pfeiffer2006,Haviland1996}.  Thereby, the  variables $\phi_n$ describing the dynamics of the superconducting islands $ $ are the  node fluxes~\cite{Supplementals,Devoret1995}, which are here the time-integrated voltages  with respect to the ground
$
	\phi_n (t) = \frac{\hbar}{2e}\int_{-\infty}^{t}dt  V_n(t).
$
Superconducting circuits allow for a large variety of realizations and a broad range of possible  parameters~\cite{Manucharyan2009,Erguel2013,Koch2009,Pfeiffer2006}.
We assume  large   $C \hbar I_{c,n}/e^2\gg 1$ and $h/4 e^2 R\gg1$, where $C$, $I_{c,n}$ and $R$ denote capacitance, critical Josephson current and resistance as depicted in Fig.~\ref{fig:sketch}(b). This parameter regime justifies to treat $\phi_n(t)$ as classical variables~\cite{Erguel2013}.
The strength of the nonlinearity can be adjusted by $\delta$~\cite{Manucharyan2009}. Additionally, the dynamics is subjected to a  monochromatic driving with amplitude $ I_{ac}$ and frequency $\Omega$. It is straightforward to derive the EoM \eqref{eq:EoM} using Kirchhoff's first law and find the relation of the physical parameters $ R, C, I_{c,n}$ and $L_{n}$ and the parameters appearing in \eqref{eq:EoM}~\cite{Devoret1995,Supplementals}.

The position-dependent couplings possess an alternating structure and read
\begin{equation}
 t_{n,n+1}= t_0- \alpha(-1)^n,
\end{equation}
where $2\alpha$ is the difference of two subsequent couplings.
Thus, the system exhibits the same coupling geometry as the  SSH model~\cite{Su1979}.

The EoM are  designed in such a way, that in the linear case $\delta=0$, the spectrum of the modes reproduce the properties of the standard  SSH model, which exhibit a topological phase transition at $\alpha=0$~\cite{Kane2014}. Thereby, the system has topologically protected-boundary modes with frequency $\omega_{b}=\sqrt{2 t_0}$ in the topologically non-trivial phase for $\alpha<0$, which are absent in the topologically trivial phase for $\alpha>0$. As we see later, features of the linear SSH model  still persist in the chaotic dynamics of the  nonlinear model.

\begin{figure}
\centerline{\includegraphics[width=\columnwidth]{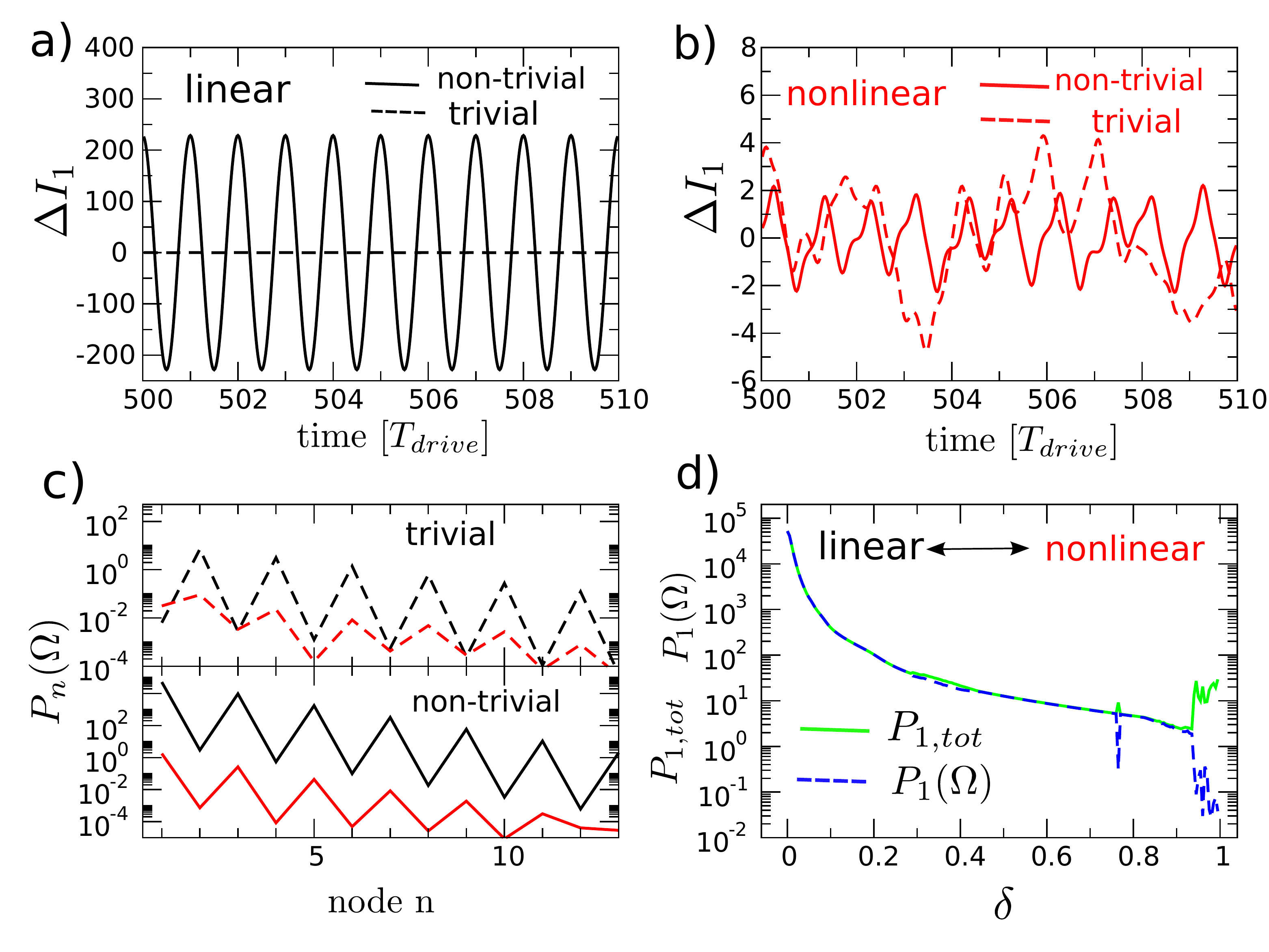}}
\caption{ (a) Current $\Delta I_1=I_1(t)- I_{bulk}(t) $ in units of $t_0/\mathcal R$ flowing from node $n=1$ through the resistance $R$ to the ground in the linear system $\delta=0$.  Parameters are $\alpha = -0.4 t_0$ (non-trivial, solid) and $\alpha = 0.2\,t_0$ (trivial, dashed), $\mathcal R = 0.02 \,t_0 \Omega$ and $N=200$. $T_{drive}=2\pi /\Omega$  denotes the driving period, where $\Omega=\sqrt{2t_0}$. (b) as in (a) but for $\delta=0.95$. In this case, the system exhibits chaos.  (c) $P_n(\Omega)$ in units of $(t_0/\mathcal R)^2$ in Eq. \eqref{eq:PowerSpectrum} for the time evolutions in (a) and (b). Throughout this letter,  we take $t_{min}=500\, T_{drive}$ and  $\tau=100 \,T_{drive}$ to evaluate  Eq.~\eqref{eq:PowerSpectrum}.  (d) Dependence of $P_1(\Omega)$ and $P_{1,tot}$ in Eq.~\eqref{eq:totalPower} as a function of $\delta$.}
\label{fig:timeEvolutions}
\end{figure}%

\textit{Time evolution.}
In Fig.~\ref{fig:timeEvolutions}(a) and (b) we depict the time evolution of  node $n=1$ for  $ \delta=0$ and $\delta=0.95$, respectively.
Throughout the Letter, we choose to drive with a frequency $\Omega= \omega_b$ corresponding to the topologically-protected boundary mode appearing for $\delta=0$ and $\alpha<0$ to elucidate the topological effects. 
Instead of depicting the node fluxes $\phi_n(t)$, we consider
\begin{equation}
I_n(t) = \mathcal R \dot \phi_n(t).
\end{equation}
This quantity is proportional to the current flowing from  node $n$ through the resistance $R$ to the ground and is therefore experimentally accessible~\cite{Erguel2013}. Additionally, we find, that $I_n$ instead of $\phi_n$ is more appropriate for our investigation, as slow contributions in $\phi_n$ have a smaller weight. 

We  always choose $\phi_n(t=0)=0$ as initial state.
In  panels (a) and (b), we show the time evolution after an initial transient phase in order to make sure that we have approached the corresponding attractor. To obtain a clearer understanding, we  depict the difference $\Delta I_n (t)=I_n(t)- I_{bulk}(t)$, where  $I_{bulk}(t)$ denotes  the bulk current. This is the time-periodic current under a periodic boundary condition $\phi_{N+1}= \phi_n$ and reads $I_{bulk}(t) = \text{Im}\, \left[\phi_0 \Omega e^{i\Omega t}\right]$ with $ \phi_0 = I_{ac}/\Omega (i \mathcal R -\Omega) $~\cite{Supplementals}.

For the parameters in panels (a), the time evolution exhibits   a harmonic oscillation. 
Due to the subtraction of the bulk current, the oscillation at node $n=1$ for $\alpha=0.2$ (trivial phase) vanishes nearly completely, while the oscillation amplitude is extremely large for $\alpha=-0.4$ (non-trivial  phase). To further analyze this dynamics, we consider the position-dependent power spectral density~\cite{Kautz1996}
\begin{align}
	\label{eq:PowerSpectrum}
P_n(\omega) &=\left| \tilde I_n(\omega)\right|^2 \qquad \text{with}\\
\tilde I_n(\omega)&= \lim_{\tau\rightarrow \infty}\frac{2}{\tau} \int_{t_{min}}^{t_{min}+ \tau }dt\left( I_{n}(t)-I_{bulk}(t)\right) e^{i \omega t}. \nonumber
\end{align}
For long times, the dynamics of the linear system  displays  harmonic oscillations with  frequency $\Omega$ of the external driving. For this reason, we depict $P_n (\Omega)$ in Fig.~\ref{fig:timeEvolutions}(c). Here we observe an alternating pattern of finite and almost zero power as a function of n. Thereby, the power is finite on odd (even) nodes in the non-trivial (trivial) phase. This is a typical topological feature of the linear model \cite{Asboth2016,Supplementals}.

For a finite $\delta$, the system can exhibit a chaotic time evolution as depicted in Fig.~\ref{fig:timeEvolutions}(b). Surprisingly, despite of the chaotic dynamics,  the power spectral density still exhibits an alternating structure. Note that the overall power is considerably smaller than in the linear case. This is a consequence of the nonlinearity, which we investigate in Fig.~\ref{fig:timeEvolutions}(d), where we depict  $P_1 (\Omega)$ and the position-resolved total power
\begin{equation}
P_{n,tot} \equiv \int_{0}^{\infty}d\omega P_n(\omega) 
\label{eq:totalPower}
\end{equation}
for $n=1$ as a function of $\delta$. We observe, that starting from $\delta=0$, the power rapidly decreases. This happens, as the driving frequency $\Omega$ is no more in resonance with the  boundary mode of the linear system, which is modified due to the nonlinearity $\delta$. For $\delta=0$,  $P_{n,tot}$ and $P_n(\Omega)$  coincide, as the time evolution is harmonic with  frequency $\Omega$. This situation can be observed for a broad range of $\delta$ values. In a region around $\delta \approx 1 $, both quantities strongly deviate and we find  chaos.  We are interested in this region, so that we concentrate on $\delta=0.95$ in the remainder of this Letter.

\begin{figure}
\centerline{\includegraphics[width=0.99\columnwidth]{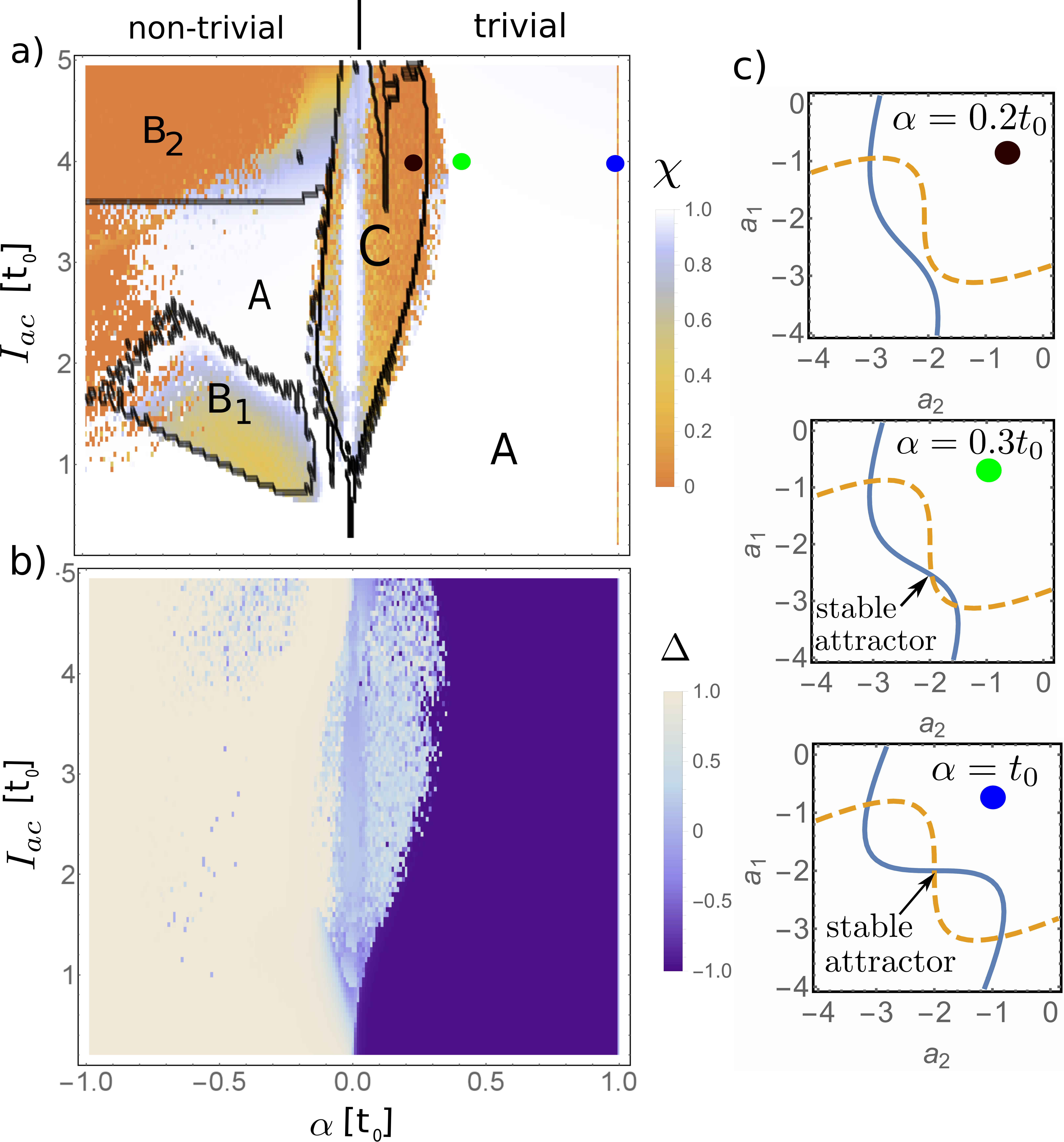}}
\caption{(a) Phase diagram for the order parameter $\chi$ defined in Eq.~\eqref{eq:DefChi}.  Parameters are as in Fig.~\ref{fig:timeEvolutions}. Black lines depict the phase boundary obtained by the generalized force functionals Eq.~\eqref{eq:effectiveEq}. (b) Phase diagram of the topological order parameter $\Delta$. (c) Level sets  $G_1=0$ (solid) and $G_2=0$ (dashed) of Eq.~\eqref{eq:effectiveEq} for fixed, but optimized $a_3$ for $ I_{ac}=4 t_0$. The corresponding parameters are marked in (a) by colored dots. The vanishing of the stable attractor at $\alpha\approx 0.25 t_0$ triggers the chaotic dynamics observed in the central region of the phase diagram. 
}
\label{fig:phaseDiagram}
\end{figure}%
%
%
\textit{Order parameter.} A useful quantity which gives insight into the dynamics of the system is given by
\begin{equation}
	\chi =  \frac{ P_{1}(\Omega) + P_{2}(\Omega)}{P_{1,tot} + P_{2,tot}},
	\label{eq:DefChi}
\end{equation}
which we introduce as an order parameter for the phase diagram in Fig.~\ref{fig:phaseDiagram}(a). There we depict $\chi$ as function of $\alpha$ and $ I_{ac}$, where we observe several regions among which we find periodic and chaotic dynamics.
If the system synchronizes with the external driving then $P_{n,tot}\approx P_n(\Omega)$ and $\chi=1$. Contrary, for  chaotic dynamics, the power distributes over many modes, so that $\chi\approx 0$, as can be seen in Fig.~\ref{fig:timeEvolutions}(d) for $\delta\approx 1$. Instead of defining $ P_{1}(\Omega)/P_{1,tot} $ as an order parameter, we choose to incorporate the power of $n=2$ in Eq.~\eqref{eq:DefChi}. In doing so, we avoid a division by very small $P_{1,tot}$ appearing, e.g., for  $\alpha\gg 0$. The regions marked by A exhibit  periodic dynamics, while in $B_1$ we observe quasiperiodic dynamics. In the regions labeled by $B_2$ and $C$ we find a chaotic time evolution.We also calculated the power spectrum and the Lyapunov exponent (not shown) to verify that the  dynamics  is indeed chaotic.

\textit{Topological character}.
The chaotic dynamics in regions $B_2$ and $C$ is qualitatively different. To see this, we consider the following quantity
\begin{equation}
\Delta =  \frac{ P_{1}(\Omega) - P_{2}(\Omega)}{P_{1}(\Omega) + P_{2}(\Omega)}.
\label{eq:DefDelta}
\end{equation}
In the linear system, $\Delta=1$ and $\Delta=-1$ in the non-trivial and trivial phase, respectively (compare with Fig.~\ref{fig:timeEvolutions}(c)). In Fig.~\ref{fig:phaseDiagram}(b), we investigate how this quantity is modified in the nonlinear system $\delta=0.95$ for increasing driving amplitude $ I_{ac}$. For small driving $ I_{ac}$, the time evolutions corresponds to the one of the linear model $\delta=0$. In this case we observe a fast crossover from  $\Delta=1$ to $\Delta=-1$ at $\alpha\approx 0$.

It is very surprising to see, that there is a clear topological character $\Delta$ in wide parts of the phase diagram. Even more appealing is the observation, that region $C$ can be clearly recognized in Fig.~\ref{fig:phaseDiagram}(b), while  region $B_2$ can not. More precisely, the underlying topology  in $B_2$ is stronger pronounced than in $C$. As we show in the next part, there is also a different mechanism behind the appearance of chaos in these two regions.

\textit{Time-independent effective equations}.
To gain more insight, we  derive time-independent nonlinear equations that capture the underlying processes.
We observe that the time evolution of $\phi_n$ in the regular regimes $\chi\approx1$ is essentially given by a harmonic oscillation up to a small correction $\Delta_n(t)$. Accordingly, we split the time evolution as~\cite{SHAPIRO1964}
\begin{equation}
	\phi_{n}(t)= a_{n} \cos(\Omega t)  +\Delta_n(t).
	\label{eq:Ansatz}
\end{equation}
The dynamics in zeroth order of $\Delta_n$ is thus determined by the amplitudes $a_{n}$. After inserting  ansatz \eqref{eq:Ansatz} into the EoM \eqref{eq:EoM}, we perform a Fourier analysis. In doing so, we obtain  a set of nonlinear equations~\cite{Supplementals}
\begin{align}
	 0&= t_{n-1,n}  F_\delta \left(a_{n-1}-a_{n} \right)+  t_{n,n+1}  F_\delta \left(a_{n+1}-a_{n} \right) \nonumber \\
  &+  I_{ac}  +\Omega^2 a_{n}\equiv G_n \left[ \left\lbrace a_n \right \rbrace\right] ,
  \label{eq:effectiveEq}
\end{align}
with
\begin{equation}
	F_\delta(x) = \left(1-\delta\right)x+  2 \delta  \mathcal J_1\left( x\right)\equiv \frac{d}{dx} V_{eff} (x),
\end{equation}
which determine the amplitudes  $a_n$. Here $\mathcal J_1(x)$ denotes the first-order Bessel function. The $G_n$ can be considered as generalized force functionals in Fourier space and $V_{eff} (x)$ as an effective coupling potential. The latter is depicted in Fig.~\ref{fig:sketch}(d).   In the derivation, we have neglected the dissipative term, as $\mathcal R$ is small. A linear stability analysis for $\Delta_n$ reveals the stability of the obtained amplitudes $a_n$. In order to distinguish phases $B_2$ and $C$, we numerically  minimize 
$$\mathcal G = \sum_n G_n^2,$$
 instead of finding a root of $G_n=0$ and  check if the minimum of $\mathcal G$ is  a root of \eqref{eq:effectiveEq}. 
As $\mathcal G$ exhibits a large number of minima, it is important to find the one corresponding to the actual steady-state dynamics.  We choose a starting point which resembles the amplitudes $a_n$ of the steady state of the linear system, up to a normalization~\cite{Supplementals}.
We find that our approach reproduces the actual dynamics with high accuracy  where $\chi\approx 1$.

\textit{Fix-point analysis}.
 The outcome of the fix-point analysis of \eqref{eq:effectiveEq} is included  in Fig.~\ref{fig:phaseDiagram}(a) by black lines. Thereby, we distinguish  three cases. First, the minimum of $\mathcal G$ discovered by the numerics is a root of \eqref{eq:effectiveEq} and stable in the linear stability analysis (region  A). Second, we discover a root, but it is linearly unstable (region  B). Third, the minimum of $\mathcal G $  is not a root of  \eqref{eq:effectiveEq} (region  C). The most interesting case is the latter as, according to the following fix-point analysis, this has a topological origin. To understand this, we first investigate the limiting cases $\alpha=\pm t_0$ in more detail.

For $\alpha= t_0$, the system consists of $N/2$ decoupled dimers, as sketched in Fig.~\ref{fig:sketch}(e). We depict the level sets of $G_1=0$ and $G_2=0$ in Fig.~\ref{fig:phaseDiagram}(c).  We observe a symmetric pair of  lines which intersect three times, thus, there are three distinct fix points, where only the middle one is a stable attractor. Altogether, the chain  thus exhibits  $3^{N/2}$ fix points for  $\alpha= t_0$. 

 In the case $\alpha=-t_0$, we have $N/2-1$ decoupled dimers and two isolated nodes at the  ends of the chain as sketched in  Fig.~\ref{fig:sketch}(f). The function $G_1$  does not depend on $a_2$.   $G_1$ has in this case only one root (this is also true for $G_N$). Altogether the chain has $3^{N/2-1}$ fix points.  Thus, there is a different number of fix points in the limiting cases  $\alpha=\pm t_0$. Consequently, when varying $\alpha$ from one limiting case to the other one, there are topologically-enforced bifurcations. In particular, as the stable fix points of the limiting cases $\alpha=\pm t_0$ are structurally different, there is no way to smoothly transform one into the other without  bifurcation.
 
 To illustrate this, we included in Fig.~\ref{fig:phaseDiagram}(c) the panels for $\alpha=0.2 t_0$ and $\alpha=0.3 t_0$. Thereby, 
 we insert $a_3$ found by the numerical minimization of $\mathcal G$ into the equation for $G_2$ as a fix parameter. The two panels depict the situation shortly before and after the bifurcation. 
 This bifurcation  is a so-called  saddle-node  bifurcation, where  two fix points  annihilate each other by varying $\alpha$~\cite{Strogatz2014}. 
 
 The middle fix point  corresponds to a stable attractor of the system. When we  lower $\alpha$, the stable attractor  vanishes in a bifurcation, and the unstable fix point remains (panel for $\alpha=0.2 t_0$). Consequently, there is  no stable periodic attractor, so that the dynamics gets chaotic. By further decreasing $\alpha$, the remaining root can either become stable so we enter again in a regular regime, or stays unstable, which finally results in the chaotic phase $B_2$.

\textit{Discussion}.
 Our investigations reveal interesting effects appearing in the nonlinearly-rendered SSH model. 
 The time evolution  exhibits period dynamics, quasiperiodicity and even chaos. By introducing the order parameter $\Delta$ quantifying the topological character of the dynamics, we found that there are two types of chaotic dynamics, only one of which is indicated clearly by $\Delta$. In the other chaotic region, the time evolution surprisingly still exhibits the topological features of the linear model. We  emphasize, that the order parameters $\Delta$ and $\chi$ are experimentally accessible by measuring the current of the first two nodes $n=1,2$ only. This could be possible with similar experimentally techniques as in Ref.~\cite{He1984,Manucharyan2009,Erguel2013,Koch2009,Pfeiffer2006,Haviland1996}.
 
 Moreover, based on a Fourier analysis of the EoM, we have identified the reason for the  chaotic region separating the two areas with distinct topological character $\Delta=\pm 1$. Comparing the structure of the fix points of the two topological limiting cases, we found that it is not possible to smoothly transform one into the other without a bifurcation. Thereby, the previously stable fix point vanishes, which gives rise to chaos. This is in strong analogy to the topology of the linear system, where the presence and absence of topologically-protected boundary modes is also apparent from a consideration of the topological limiting cases. Despite of this analogy, it is not possible to apply the topological concepts known from the linear model, namely the winding number~\cite{Asboth2016}, to describe the topology of our nonlinear model, which refers to a fix point analysis. Nevertheless, the topological-enforced bifurcation and the topological phase transition of the linear model are both independent of the system size due to the previously mentioned arguments,
 which we confirmed by simulating smaller system sizes (not shown). For instance for  20 nodes, the  phase diagram Fig.~\ref{fig:phaseDiagram}(a) exhibits larger chaotic regions in the nontrivial part $\alpha<0$.  
 We also mention that the topologically-induced chaos is reminiscent to the topological instability appearing at the  phase transition considered in Ref.~\cite{Engelhardt2016}, although the underlying reason is different.
 
 Finally, we emphasize that due to their topological origin, our findings do not depend on   details of the system. The topological-enforced bifurcations appear also, e.g.,  with different kind of dissipation or for $\delta=1$. The latter case is particularly important as such kind of Josephson junction arrays are used to fix the voltage standard~\cite{Hamilton2000}. So this kind of setup could also be used to test our findings. Furthermore, even the form of the nonlinearity is not relevant. Bifurcations even occur for, e.g., a $-x^3$ term in Eq.~\eqref{eq:couplingFunction} instead of the sine, which also suggest that our findings can also appear in other kind of systems.
  We also suppose that the effects discussed here appear in more complex system with an underlying topological coupling geometry, as in two-dimensional nonlinearly-rendered topologically arrays, like in nonlinear versions of the Hofstadter or Haldane models~\cite{Hofstadter1976,Haldane1988}

\textit{Acknowledgments}
The authors gratefully acknowledge financial support
from the DFG Grants No. BR 1528/7, No. BR 1528/8,
No. BR 1528/9, No. SFB 910 and No. GRK 1558 as well as inspiring discussion with Jordi Pic\'o, Jan Totz and Anna Zakharova.
 This work was supported by the Spanish Ministry
through Grant No. MAT2014-58241-P.

\bibliographystyle{phaip}
\bibliography{topology}

\newpage

\begin{widetext}

\begin{center}
   \huge \textbf{Supplementary Information}
   \vspace{0.7cm}

\end{center}

  \setcounter{section}{0}

\section{ Equations of motion in a superconducting circuit} 
  
  \label{app:equations}
  Kirchhoff's first law states that all currents flowing into a node sum up to zero. For the circuit  in Fig.~1(b) in the Letter this means
  \begin{equation}
  \mathcal I_{n}+\mathcal I_{n;C}+ \mathcal I_{n;-}+ \mathcal I_{n;+}=0,
  \label{eq:KirchoffLaw}
  \end{equation}
  where $I_n$ and  $I_{n;C} $ denote the current flowing through the resistance and the capacity, respectively~\cite{Devoret1995}.
  Expressing them with the fluxes  $\phi_{n}$, we find
  \begin{equation}
  \mathcal I_{n} = \frac 1{R} V_{n} =  \frac 1{R} \frac {\hbar}{2e} \dot \phi_{n} 
  \end{equation}
 and 
  \begin{equation}
  \mathcal I_{n;C} = C \left(\dot  V_{n} - \dot U_{ext}(t) \right)=  C   \frac {\hbar}{2e } \ddot  \phi_{n}  - \mathcal I_{ext}(t),
  \end{equation}
  where  $V_{n}$ is the voltage of  node $n$ with regard to the ground. Additionally, we have defined the external current $I_{ext}(t)$. The currents coming from node $n\pm1$  denoted as $ I_{n;\pm}$ read
  \begin{equation}
  \mathcal I_{n;\pm} = I_{c,\eta_\pm}\sin\left( \phi_{n\pm 1}-\phi_{n} \right)+  \frac {\hbar}{2e} \frac 1{L_{\eta_\pm}} \left( \phi_{n\pm 1} -  \phi_{n} \right),
  \end{equation}
  where $\eta_\pm =n \mod 2+ \delta_{1,\mp 1 }$.
 Inserting this into Eq.~\eqref{eq:KirchoffLaw} and resolving for $\ddot \phi_n$, we obtain Eq.~(1) in the Letter  using the  definitions
  \begin{align}
\frac { I_{c,\eta_\pm}}{C} &\equiv (t_0 - \alpha (-1)^n) \delta  ,\qquad
  &\frac {\hbar}{2e} \frac 1{C L_{\eta_\pm}   } &\equiv   (t_0 - \alpha (-1)^n)  (1-\delta),\\
 \frac 1{C R} \frac {\hbar}{2e}   &\equiv \mathcal R ,
&\frac{\mathcal I_{ext}(t)}{C}  &\equiv   I_{ext}(t).
  \end{align}
Finally, we choose the external driving to be
\begin{equation}
	I_{ext}(t) = I_{ac} \cos \Omega t.
\end{equation}

\section{Effective time-independent equations}

Here we provide more details concerning the derivation of the effective equations and the calculation of the reconstructed phase diagram. We insert the ansatz Eq.~(9) in the Letter into the equations of motion (EoM) (1) and expand them up to first order in $\Delta_n$. By expanding the appearing $\sin\left[(a_{n+1}-a_n)\cos(\Omega t)  \right]]$ in a Fourier series in terms of Bessel functions, we obtain \cite{SHAPIRO1964}
\begin{align}
-\Omega^2 a_{n}\cos \Omega t + \ddot \Delta_n &=  t_{n,n+1}(1-\delta) (a_{n+1}-a_{n})\cos(\Omega t)  + t_{n,n+1} \delta 2  \sum_{m=0}^{\infty} \mathcal J_{2m+1}(a_{n+1}-a_{n}) \cos\left[ (2m +1) \Omega t \right] \nonumber  \\
 &+t_{n,n-1}(1-\delta) \left[(a_{n-1}-a_{n})\cos(\Omega t)  \right] + t_{n,n-1} \delta 2 \sum_{m=0}^{\infty} \mathcal J_{2m+1}(a_{n-1}-a_{n}) \cos\left[ (2m +1) \Omega t \right]  \nonumber \\
 &+\Omega \mathcal R a_{n} \sin (\Omega t)+ I_{ac}   \cos(\Omega t)  \nonumber \\
 &+ t_{n,n+1}(1-\delta) (\Delta_{n+1}-\Delta_{n}) + t_{n,n+1} \delta \cos\left[(a_{n+1}-a_{n}) \cos(\Omega t)\right](\Delta_{
 n+1}-\Delta_{n}) \nonumber  \\
   &+t_{n,n-1}(1-\delta) (\Delta_{n-1}-\Delta_{n}) + t_{n,n-1} \delta \cos\left[(a_{n-1}-a_{n}) \cos(\Omega t\right](\Delta_{n-1}-\Delta_{n}) \nonumber  \\
   &- \mathcal R  \dot  \Delta_n,
   \label{eq:ExpandedEquation}
\end{align}
which is exact up to first order in $\Delta_n$.

As explained in the Letter, we omit here the term proportional to $\sin(\Omega t)$ as this has anyway a minor influence on the dynamics since $\mathcal R$ is small. 
We require that all terms proportional to $\cos(\Omega t)$ vanish, which gives us  Eq.~(10). Assuming that $\phi_n(t)$ and $a_n \cos(\Omega t)$ in ansatz Eq.~(9) in the Letter  are both solutions of the EoM, we finally get the EoM for the deviations

\begin{align}
	\ddot \Delta_n &=  t_{n,n+1}(1-\delta) (\Delta_{n+1}-\Delta_{n}) + t_{n,n+1} \delta \cos\left[(a_{n+1}-a_{n}) \cos(\Omega t)\right](\Delta_{n+1}-\Delta_{n}) \nonumber \\
	   &+t_{n,n-1}(1-\delta) (\Delta_{n-1}-\Delta_{n}) + t_{n,n-1} \delta \cos\left[(a_{n-1}-a_{n}) \cos(\Omega t)\right](\Delta_{n-1}-\Delta_{n})     .
\end{align}
In this differential equation the variables $\Delta_n$ appear only linearly.  The terms proportional to $\delta$ in the  first and second line constitute a periodic driving. This driving can lead to an exponential growth of the variables $\Delta_n$ as a function of time. In the reconstruction phase diagram, we thereby denote the root of $ G_n=0$ in Eq.~(10) to be unstable, if the time evolution of $\Delta_n$ exhibits a continuing growth for the initial condition $\phi_n=1$ for even $n$ and $\phi_n=-1$ for odd $n$.

\section{Steady state of the linear system}

In this section, we derive an exact expression for the periodic dynamics of the linear system for $\delta=0$ in the long-time limit. 
For the sake of simplicity,  we consider the semi-infinite system, so that we can resort to the so-called transfer-matrix method. To enable a better analytical treatment,  we complexify the EoM by replacing $ \cos\Omega t$ by $ e^{i\Omega t} $ and use the ansatz $\phi_{n}=\phi_{n,0} e^{i\Omega t} $. In doing so, the EoM get time independent and read
\begin{align}
	-\Omega^2 \phi_{n,0} =
	t_{n,n+1}   \left( \phi_{n+1,0}-\phi_{n,0} \right) + t_{n,n-1}    \left( \phi_{n-1,0}-\phi_{n,0} \right) 
	 - i \Omega \mathcal R  \phi_{n,0}  + I_{ac}.
\end{align}
In order to  get rid of the inhomogeneity, we transform the  equation  by
\begin{align}
	 \phi_{n,0}- \phi_0 =
	\tilde t_1   \left( \phi_{n+1,0}-\phi_0-\phi_{n,0}+\phi_0 \right) +\tilde  t_2   \left( \phi_{n-1,0}-\phi_0-\phi_{n,0} +\phi_0\right) ,
\end{align}
where $\phi_0$ is the solution of the translationally-invariant system with periodic boundary condition $\phi_{n+1}=\phi_1$, which reads
\begin{equation}
	\phi_0 = \frac{I_{ac}}{\Omega (i \mathcal R-\Omega)}
	\label{eq:amplitudeTranslationalInvariant}
\end{equation}
 and 
\begin{equation}
\tilde t_{n,n+1} = \frac{t_{n,n+1}}{\Omega(i \mathcal R - \Omega)}.
\end{equation}
Finally we define the new coordinates $\xi_{n,0}=\phi_{n,0}- \phi_0  $, so that the equation to solve now reads
\begin{align}
	 \xi_{n,0} =
	\tilde t_{n,n+1}   \left( \xi_{n+10}-\xi_{n,0} \right) +\tilde  t_{n,n-1}  \left( \xi_{n-1,0}-\xi_{n,0}\right) 
\end{align}
This is now in an appropriate form to apply the transfer-matrix method. To propagate the mode function from one node $n$ to the next one $n+1$, we can use the relation
\begin{equation}
\left( \begin{array}{c}
\xi_{n+1,0} \\
\xi_{n,0}
\end{array}
\right) =
\mathbf M_n
\left( \begin{array}{c}
\xi_{n,0} \\
\xi_{n-1,0}
\end{array}
\right) 
\qquad \text{where}
\qquad
\mathbf M_n=
\left( \begin{array}{cc}
 \frac {A}{\tilde t_{n,n+1} \Omega(i \mathcal R -\Omega)} & -\frac {\tilde t_{n,n-1}}{\tilde t_{n,n+1}}\\
 1  & 0
\end{array}
\right) 
\end{equation}
and $A= (t_{n,n+1}+ t_{n,n-1})+\Omega(i \mathcal R -\Omega)$. 
 Defining $t_1\equiv t_0+\alpha$ and $t_2\equiv t_0-\alpha$, it is not hard to see that for even $n$
\begin{equation}
\left( \begin{array}{c}
\xi_{n+1,0} \\
\xi_{n,0}
\end{array}
\right) =
\left(\mathbf M\right)^{n/2}
\left( \begin{array}{c}
\xi_{1,0} \\
-\phi_0
\end{array}
\right) ,
\end{equation}
where
\begin{equation}
	\mathbf M=\mathbf M_2 \mathbf M_1=
	\left( \begin{array}{cc}
	 \frac{A^2}{t_{1} t_{2}} - \frac{t_1}{t_2}  & -\frac{A }{t_{1}}  \\
	\frac{A }{t_{1}}   & -\frac{t_{2}}{t_{1}}
	\end{array}
	\right) .
\end{equation}

The two eigenvalues $\lambda_j$ and eigenstates of $\mathbf M $ contain the information how the wave function propagates within the bulk. In order to find a physically reasonable state, the important eigenvalue is the one with $\left|\lambda_j \right|<1$. In the topological non-trivial phase $\alpha<0 $ or  $t_1<t_2$, we therefore find
\begin{align}
\lambda_j &= \frac{1}{2 t_1 t_2 } \left(  A^2 -t_1^2 -t_2^2 + \mathcal D \right) \rightarrow -\frac{t_1}{t_2} \quad \text{for} \quad A\rightarrow 0, \\
\xi_{1,0} &=  \frac{-\phi_0}{2 A t_2 } \left(  A^2 -t_1^2 +t_2^2 + \mathcal D \right) \rightarrow   -\phi_0 \frac{t_2}{ A} \quad \text{for} \quad A\rightarrow 0,
\end{align}
where
\begin{equation}
\mathcal D = \sqrt{(A^2-t_1^2-t_2^2)^2-4 t_1^2 t_2^2}.
\end{equation}
Approximately, the  steady-state  thus reads
\begin{equation}
\phi_{n,0} \approx\phi_0+ \frac{1- (-1)^n }{2}  \left(- \frac{t_1}{t_2} \right)^{n/2} \xi_{1,0}\qquad\text{where} \quad \xi_{1,0} =-\phi_0 \frac{t_2}{A}.
\label{eq:SteadyStateLinearTopological}
\end{equation}

In the trivial phase $t_1>t_2$, we find
\begin{align}
\lambda_j &= \frac{1}{2 t_1 t_2 } \left(  A^2 -t_1^2 -t_2^2 - \mathcal D \right) \rightarrow -\frac{t_2}{t_1} \quad \text{for} \quad A\rightarrow 0 ,\\
\xi_{1,0} &=  \frac{-\phi_0}{2 A t_2 } \left(  A^2 -t_1^2 +t_2^2 - \mathcal D \right) \rightarrow  0 \quad \text{for} \quad A\rightarrow 0.
\end{align}
Thus, the steady-state amplitude in the trivial phase is approximately
\begin{equation}
\phi_{n,0} \approx\phi_0 - \frac{1- (-1)^{n+1}}{2} \left(- \frac{t_2}{t_1} \right)^{n/2}  \phi_{0} .
\label{eq:SteadyStateLinearTrivial}
\end{equation}

\section{Initial values of the minimization procedure}

We aim to find a starting point, so that the minimum of $\mathcal G$ found by the numerics resembles the amplitudes of the steady state in the regular regions of the phase diagram. Therefore, the starting point should be already quite close to the actual minimum of $\mathcal G$ corresponding to the steady-state dynamics.

The steady-state dynamics is very similar to the topological boundary excitation of the linear system  presented in the previous section. For this reason, we take the steady-state of the linear system as given in Eqs.~\eqref{eq:SteadyStateLinearTopological} and \eqref{eq:SteadyStateLinearTrivial} as initial value for the minimization, but renormalize the  overall oscillation amplitude. More precisely, the starting point amplitudes $a_n^0$ of the minimization shall fulfill
\begin{equation}
\frac{ \phi_{n,0}-\phi_{0} }{ \phi_{n',0} - \phi_{0}}= \frac{a_n^0-\phi_{0}}{a_{n'}^0-\phi_{0}}.
\label{eq:ratioCondition}
\end{equation}
 As the amplitudes $\phi_{n,0}$ are structurally different in both topological phases, we treat the two cases independently.
 In the non-trivial phase, we start at node $n=1$. Having found an approximate expression for $ a_{1}^0$, all other amplitudes  $ a_{n}^0$ can be easily determined by relation~\eqref{eq:ratioCondition}. In the numerics we observed that $ a_1-\phi_{0}$ only slightly depends on $\alpha$. For this reason, we use the  limiting case $\alpha=- t_0$ to get $a_1^0$. In doing so, Eq.~(10) for $n=1$ decouples and  we can  solve it numerically to find $a_1^0$. Thereby, the root is unique.
 
 In the topologically trivial phase we start at   $ a_2^0$   as $\phi_{1,0}\approx\phi_0$. To obtain a staring value for  $ a_2^0$, we use that $ a_1\approx \phi_0$ in the numerics. We again use (10) for $n=1$ but with $ a_1\rightarrow \phi_0$. After having found a starting value for $a_2^0$, we find all other $a_n^0$ using~\eqref{eq:ratioCondition}.

\end{widetext}

\end{document}